\documentclass[aps,twocolumn,prl,floatfix,showpacs]{revtex4}
\usepackage{graphicx}
\usepackage{amsfonts,amsmath}

\begin{document}
\title{Experimental Observation of Quantum Reflection in the High Energy Limit}
\author{V. Druzhinina}
\author{M. DeKieviet}
\affiliation{Physikalisches Institut, Heidelberg University, 69120  Heidelberg, Germany}
\date{}
\begin{abstract}
We present first experimental data on the high energy behavior of helium atoms quantum reflecting from the nanoscopically disordered surface of an $\alpha$-quartz crystal. 
The use of the light, stable and inert He atom not only opens the unique possibility of measuring quantum reflectivity in the thusfar inaccessible limit of high energies, but also allows the determination of the gas-solid interaction potential.
The specularly reflected intensity from the rough surface shows a change of 5 orders of magnitude within an incident angular range of less than $6^\circ$. 
By separating out the influence of surface disorder the quantum reflection coefficient for the smooth surface is deduced. 
Firstly, the data confirm the high energy asymptotic behavior of the reflection, defined by the non-retarded attractive van der Waals potential $-C_3/r^{3}$. The experiment shows very good agreement with our calculations covering the entire energy region, in which also Casimir forces play a role. Parameters for the gas-solid interaction perfectly match those reported in literature in the vicinity of the potential minimum.
\end{abstract}
\maketitle
Surprising quantum phenomena may occur when the wave nature of an atom becomes dominant with respect to its classical, particle-like behavior.An example is the above-barrier reflection of a slow atom, which' kinetic energy exceeds the barrier height~\cite{Pokrovskii1, Maitra}. Interestingly, for this to happen the barrier does not necessarily need to be repulsive~\cite{Aspect}. 
In the quantum regime, reflection also takes place from a purely attractive potential, which falls off with distance $r$ faster than $r^{-2}$. This was predicted in~\cite{Carraro,Cote,Friedrich} for the attractive interaction potential between a neutral atom and the surface of a solid and recently elegantly demonstrated in an experiment by Shimizu~\cite{Shimizu1}. It is not only of fundamental interest, but also has the practical prospective of using the surface of a solid as an efficient mirror for ultracold atoms. Quantum mechanically, the reflection coefficient, defined as the ratio between the reflected beam intensity and the incident one, grows to unity in the limit of very low impinging energy. 

In earlier theoretical works, Cole and Brenig~\cite{Carraro,Boeheim} establish an extremely low critical normal kinetic energy required for observing quantum reflection from surfaces, on the order of $10^{-6}$ of the interaction potential depth.
Their value is based on a coarse criterion for the break down of the WKB approximation ( $|d\lambda_{dB} (r)/dr| \approx 1$ ) and therefore represents merely a rough and rather conservative estimate of the critical energy. 
Shimizu's data, the only relevant experimental data on quantum reflection (QR) so far, were obtained in an energy regime below this and are thus no test for this criterion.
A more accurate theory on quantum reflection, developed by C{\^o}t{\'e} and Friedrich~\cite{Cote,Friedrich}, predicts a much more gradual transition from the classical into the quantum regime.\\ In this Letter we present the first experimental evidence for QR in a high energy range (up to $ 10^{-3}$ of the interaction potential well depth), therewith mapping out this transition precisely.

In addition, the exact gas-surface potential parameters are deduced, providing information on the surface properties. The exact location within the interaction potential at which QR takes place strongly depends on the kinetic energy of the incident particle. The center of this reflection region is situated approximately at some distance $r_0$, where the kinetic energy equals the potential one~\cite{Friedrich, Boeheim}. In order to observe reflection from the attractive branch of a hard wall potential it follows that the kinetic energy of the atom must be smaller than the potential depth $V_{0}$. For realistic systems there is an additional effect: soft repulsive interaction alters the form of the well near the minimum to be flatter than $r^{-3}$, which leads to a decrease of the critical energy below $V_{0}$. When reducing the incident kinetic energy, the reflection region moves towards larger distances, i.e. away from the surface, and the reflection amplitude increases. By varying the incident energy the entire attractive part of the interaction potential range can thus be sensitively probed.

In the limit of high impinging energies (close to $V_{0}$) the reflection coefficient shows asymptotic behavior. This was analytically calculated by Pokrovskii et al.~\cite{Pokrovskii1, Pokrovskii2} for above-barrier reflection in the general case, and by Friedrich et al. for the attractive atom-surface interaction~\cite{Friedrich}. The reflection coefficient from a homogeneous attractive potential $-C_n / r^{n}=-\frac{\hbar^{2}}{2m} \times (\beta_{n})^{n-2} /r^{n}$, which is a function of the distance $r$ from the surface, takes the asymptotic form:
\begin{equation}
\label{Eq:asimptota_n}
R_{n}^{asymp} =\exp[-2 \cdot B_n \cdot (k_{i} \beta _n)^{1-\frac{2}{n}}],
\end{equation}
where the constant $B_n$ depends on the power $n$ of the potential. The primary normal kinetic energy $E_i$ of the incident atom with mass $m$ and the strength of the homogeneous potential are expressed in terms of the wave number $k_i=((2m/\hbar^2)\times E_i)^{1/2}$ and the length parameter $\beta _n = ((2m/\hbar^2)\times C_n)^{1/(n-2)}$, respectively, so that the product $k_i \beta_n$ is dimensionless. The parameters of the reflecting potential can thus be determined by measuring the asymptote~(\ref{Eq:asimptota_n}) as a function of the normal incident kinetic energy.

The long range attractive part of the interaction potential between a neutral atom and the surface of a solid is in general not homogeneous, but predicted to be well described by the Casimir-van der Waals potential~\cite{Shimizu1,Friedrich}: 
\begin{equation}
\label{Eq:potential}
V(r)=-\frac{C_4}{r^{3} (r+l)},
\end{equation}
where $l$ is the transition length between the two homogeneous parts of the potential: the van der Waals potential $-C_3 / r^3 $ at the distance $r \ll l$, and the retarded potential $-C_4 / r^4 = -C_3 l/ r^4$ at $r \gg l$ due to the Casimir effect~\cite{Casimir}.

This inhomogeneous potential yields two separate reflection coefficient asymptotes of the form~(\ref{Eq:asimptota_n}), determined by the $r^{-4}$ and $r^{-3}$ parts, each one having its own validity range:
\begin{equation}
\begin{cases}
R_{4}^{asymp}(k_i) & \text{ for $G_4 \ll k_i \beta_4 \ll \rho ^2$}, 
\hfill(\number\value{equation})\\
\newcounter{asimptota_4}
\setcounter{asimptota_4}{\value{equation}}
\addtocounter{equation}{1}
R_{3}^{asymp}(k_i) & \text{ for $ \rho ^3 \ll k_i \beta_3 < (\frac{\beta_3}{a})^{\frac{3}{2}}$}.
\hspace{2.2cm}(\number\value{equation})
\newcounter{asimptota_3}
\setcounter{asimptota_3}{\value{equation}}
\addtocounter{equation}{1}
\end{cases}
\nonumber
\end{equation}
The distance $a$ denotes the position of the potential minimum and the constants $B_n$, which enter in~(\ref{Eq:asimptota_n}), are calculated to amount $B_4=1.69443$ and $B_3=2.24050$. The lower limit $G_4=0.35$~\footnote{$G_4$, given for a general case in~\cite{Pokrovskii1}, is calculated here explicitly for atom-surface interactions. In~\cite{Friedrich} this lower limit was taken to be unity.} for $k_i\beta_4$ defines the region of high energies, $E_i \gg (2m/\hbar^2)^2\times G_4/C_4$, where the reflection coefficient takes the analytic form (\ref{Eq:asimptota_n}). 

The dimensionless parameter
\begin{equation}
\label{Eq:rho}
\rho =\frac{\beta_3}{\beta_4}=\sqrt{\frac{2m}{\hbar ^{2} }} \cdot \frac{C_3 }{\sqrt{C_4}}
\end{equation}
is characteristic for the atom-surface system in general and determines the asymptotic behavior of the reflectivity: at $\rho^2 \approx \beta_3/a \gg G_4$ only the asymptote~(\number\value{asimptota_4}) can be observed, whereas when $\rho^2\not\gg G_4$ the asymptotic behavior~(\number\value{asimptota_3}) dominates. 

In order to measure the asymptote~(\number\value{asimptota_3}), determined by the non-retarded van der Waals potential only, the incident atom should have a normal incident energy $E_i \gg C_3^4/C_4^3=C_3/l^3$. 
In the entire validity range of~(\number\value{asimptota_4}) and (\number\value{asimptota_3}) for high energy QR, the incident atom is reflected relatively close to the surface. In the recent experiment by Shimizu~\cite{Shimizu1} exactly this asymptotic region was not accessible, because the metastable Ne atoms they used decay at the distance of some nm from the surface.  

In this Letter, we report on the first observation of QR of neutral helium atoms from an $\alpha$-quartz crystaline surface in the high energy limit. In this system the asymptotic behavior of the reflection coefficient is determined by the non-retarded van der Waals potential only.

The experimental results presented here are obtained on an apparatus designed for surface studies, using the novel atomic beam spin echo technique~\cite{MDeKieviet:prl95}. In this machine, the nuclear magnetic moments of $^{3}$He atoms are manipulated, so as to obtain detailed information on changes in the particle's energy before and after scattering~\cite{MDeKieviet:ss97}. For the data here, however, the actual spin echo part of the $^{3}$He spectrometer is of importance only in as much as it allows us to determine the velocity distribution in the beam precisely. 
The atomic $^{3}$He beam is produced in a 500 $\mu$m diameter nozzle source, cooled by a 4.2 K $^{4}$He bath cryostat, and detected in a commercial mass spectrometer with a saturation rate of 2 MHz. The target crystal is mounted in the scattering chamber, half way between source and detector, and can be manipulated around the 3 Cartesian axes for incident angle $\theta_i$, in-plane and azimuthal orientation. The detector can be rotated in the horizontal plane to include a total scattering angle $90^{\circ} \le (\theta_i + \theta_f) \le 180^{\circ}$ with the incident beam~\footnote{Incident ($\theta_i$) and reflected ($\theta_f$) angles are measured with respect to the surface normal.}. Since the rotation axes of incident and scattering angle are aligned to coincide, the specularly reflected He atoms can be followed directly in a so-called ($\theta-2\theta$)-scan, with an angular resolution of $\Delta\theta_f\approx 0.17^{\circ}$. Further details on the $^{3}$He spectrometer will be presented elsewhere~\cite{MDeKieviet:rsi2002}.

The QR experiment is performed on an $\alpha$-quartz single crystal having a diameter of $25$ mm, a thickness of $1$ mm and a polish on both sides. The $^{3}$He-beam average kinetic energy $E_0 = 0.63 $~meV amounts to approximately 10~\% of the He-quartz interaction potential well depth $V_{0}=9.6$~meV, reported in literature~\cite{Kunc}. The atomic beam has a wavelength distribution with a relative width of circa 20 $\%$ at an average de Broglie wavelength of 6 \AA.
AFM measurements, performed prior to chemical etching of the quartz sample, indicate a randomly stepped surface structure. The terrace width is of the order of hundred nm and their height is Gaussian distributed with width $\approx 12$~\AA. Because of an atomic roughness within the terraces, there is no specular reflection from the repulsive potential wall. Indeed, when scattering electrons (LEED), $^{4}$He or $^{3}$He atoms close to normal incidence from the surface, no reflectivity could be detected. However, upon incrementing $\theta_{i}$ beyond $84^{\circ}$ a rapidly growing $^{3}$He specular intensity is measured. 
By increasing the impinging angle, the incident kinetic energy of the atom perpendicular to the surface $E_i=E_0 \cos ^2 \theta_i$ is decreased. For $\theta_i$ ranging from $84^{\circ}$ to $89.73^{\circ}$, this means a reduction of the average normal energy from $6.9~\mu$eV down to 14 neV, corresponding to $10^{-3}$, respectively $10^{-6}$ of $V_0$. The angular width of the reflected peak at this grazing incidence is machine limited.
No broadening of the specular peak, as measured for classical reflection from stepped surfaces~\cite{Scoles}, is observed. The coherence length (or transfer width) for specular reflection, $\omega =\lambda_{dB} / (\Delta \theta _i \times \cos\theta _i )$, is 0.2 $\mu$m at normal incidence, and ranges from 2 $\mu$m up to 42 $\mu$m for the angular range in which QR is measured. The surface area illuminated by the atomic beam and the fraction of atoms actually involved in the scattering experiment depend on the incident angle. This was determined in an independent measurement and taken into consideration when analyzing the data. 
\begin{figure}
\center
\includegraphics[width=8 cm]{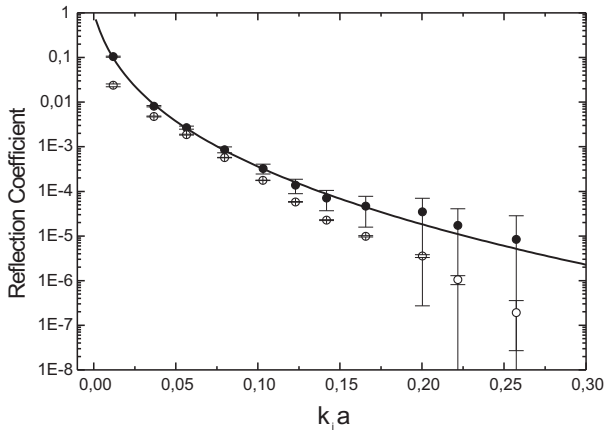}
\caption{Reflection coefficient as a function of the dimensionless average incident wave number $k_{i}a$ (proportional to $\cos \theta _{i}$).
Open circles: Experimental data from the stepped surface. Full circles: Corrected data, representing QR from the smooth surface.
Solid line: Computer simulation using the potential~(\ref{Eq:potential}) with the parameters~(\ref{Eq:potentialparameter}) and~(\ref{Eq:l}).}
\label{fig:Reflection}
\end{figure}

Fig.~\ref{fig:Reflection} shows the resulting reflection coefficient as a function of the dimensionless average normal wave number $k_{i}a=2\pi \cos\theta_i (a /\lambda_{dB})$, with $a=2.65$~\AA~being the position of the potential minimum~\cite{Kunc}. For a constant energy beam, the normal wave number is varied by changing the incident angle $\theta_i$. Open circles in the figure represent the experimental data from the randomly stepped surface. Since the step height distribution is Gaussian with width $\sigma \alt 12$ \AA, the terraces are wide and the illuminated surface area consists of a large number of them, the reflection coefficient of the rough surface (open circles) can be related to that of a smooth one (full circles), through 
\begin{equation}
\label{Eq:Rough}
R_{rough} (k_i) = e^{-4\sigma^{2}k_{i}^{2}} \cdot\frac{\int_{0^{\circ}}^{\theta} f(\theta)\,d\theta }{\int_{0^{\circ}}^{90^{\circ}} f(\theta)\,d\theta}\cdot R_{smooth}(k_i).
\end{equation}
The first term quantifies the reduction of the reflection coefficient due to dephasing of the wave function upon scattering from terraces at different heights. This effect is more pronounced at higher energy, because then the de Broglie wavelength normal to the surface, $\lambda_{dB}/\cos\theta_i$, becomes comparable to $\sigma$.
The second factor in~(\ref{Eq:Rough}) takes into account the loss of atoms hitting the steps from the side and becomes noticeable only at grazing incidence. Here, $\theta=(90^{\circ}-\theta_i)$ and $f(\theta)=(L/\sigma)/\sqrt{2\pi}\times \exp(-(L\tan{\theta}/\sigma)^2/2)$ describes the probability, that a step has height $L\tan{\theta}$. In Fig.~\ref{fig:Reflection}, the average terrace width $L$ is taken to be $L=75$~nm and $\sigma=(10 \pm 2)$ \AA.\\ The error bars on the corrected data (full circles) contain both the statistical experimental error and the uncertainty in $\sigma$.  

We can now directly compare the corrected experimental data with a computer simulation for QR from a smooth surface (solid line). Our calculation is based on the method suggested in~\cite{Cote} for an attractive potential of the form~(\ref{Eq:potential}) and shows very good agreement with the data. This method matches the WKB wave function to the exact solution of the Schr\"odinger equation in every point of the interaction. We have seen no significant difference when including the entire wavelength distribution (as determined using the spin echo technique) in the simulation instead of just the average value of $\lambda_{dB}$.  

The $C_4$ coefficient for the inhomogeneous interaction potential (\ref{Eq:potential}) entering into the computer simulation, can be written as~\cite{Dz,Shimizu1}       
\begin{equation}
\label{Eq:potentialparameter}
C_4 = \frac{1}{4\pi \varepsilon _0 } \cdot \frac{3\hbar c \alpha }{8\pi } \cdot \phi (\varepsilon )\cdot \frac{\varepsilon -1}{\varepsilon +1}
= 23.6~{\rm eV\AA}^4.
\end{equation}
Herein, $\alpha = 2.3 \times 10^{-41}$ Fm$^{2}$ denotes the polarizability of the incident He atom and $\varepsilon =4.5$ is the dielectric constant of the $\alpha$-quartz crystal~\cite{Tabele}.
 The terms containing $\varepsilon$ in expression~(\ref{Eq:potentialparameter}) correct the interaction with a dielectric surface for that with a conductive one, whereby $\phi (\varepsilon )$ is found in~\cite{Dz}. 
The transition length $l$, the only adjustable parameter in the simulation, is determined to be 
\begin{equation}
\label{Eq:l} 
l = (10 \pm 1)~{\rm nm}
\end{equation}
in order to give best agreement with experiment. This value is in perfect agreement with the wavelength $\lambda/(2\pi)=9.4$~nm corresponding to the atomic transition between the electronic ground and the first excited state in helium. In addition, the important dimensionless parameter $\rho$ characterizing our system and defined in (\ref{Eq:rho}) then becomes
\begin{equation}
\label{Eq:rhochislo}
\rho = 1.9 \pm 0.2.
\end{equation}
$\rho$ being so small, the asymptotic behavior of the reflection coefficient is expected to be determined entirely by the non-retarded interaction potential $-C_{3}/r^{3}$. That is, from the two high energy asymptotes only the higher one~(\number\value{asimptota_3}) should be observed. For our system this lies at incident energies $E_i\gg 7~10^{-6}\times V_0=69$~neV, corresponding to incident angles $\theta_i\ll 89.40^{\circ}$. 
Our experimental data acquired at non-grazing incidence are expected to get very close to this asymptote.
\begin{figure}
\center
\includegraphics[width=8 cm]{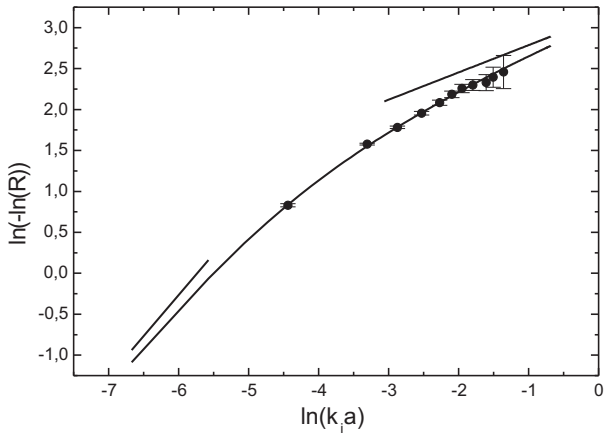}
\caption{ Reflection coefficient on a $\ln(-\ln)$-scale in dependence on $\ln(k_{i}a)$. Full circles and solid line correspond to those in Fig.~\ref{fig:Reflection}. The straight solid line of the slope $1/3$ shows the high energy asymptote~(\number\value{asimptota_3}) with $\beta _3 = 347$ \AA. The straight solid line at small $\ln(k_{i}a)$, which has slope one and the ordinate axis intercept $\approx \ln(2.4\beta _{3}/a)$ for $\rho \approx 1.9$~\cite{Friedrich}, represents the low energy asymptote.}
\label{fig:ln(-lnR)}
\end{figure}

Replotting the reflection coefficient on a $\ln(-\ln)$-scale as a function of $\ln(k_i a)$ turns the asymptotic behavior~(\ref{Eq:asimptota_n}) into a straight line, as shown in Fig.~\ref{fig:ln(-lnR)}. From $\ln(-\ln(R)) = \ln(2B_{n}(\beta_{n}/a)^{1-2/n}) + (1-2/n)\times\ln(k_{i} a)$ full information on the homogeneous part of the reflecting potential can be obtained: the slope gives the potential power $n$ and the ordinate axis intercept yields the length parameter $\beta_{n}$ (and therewith $C_{n}$).\\ The high energy asymptote for the non-retarded branch ($n=3$) in Fig.~\ref{fig:ln(-lnR)} results in a van der Waals coefficient $C_{3}=236$~meV \AA$^{3}$. 

Kunc et al.~\cite{Kunc} calculate the potential power $n$ to vary from 3.8 to 6 within the distance $r < 50$ \AA. Their potential follows (\ref{Eq:potential}) with the given parameters only at the distances $\alt 10$ \AA~from the surface. This is perfectly consistent with the authors not including retardation in their model. As an independent check of the resulting attractive potential the parameters for the potential minimum with and without retardation are calculated. The latter show agreement within $3\%$ with the values given in~\cite{Kunc}.  

In contrast, potential parameters can principally not be derived from the near-threshold $E_i\rightarrow 0$ asymptote~\cite{Friedrich} on a ln(-ln)-scale (the lower-left straight line in Fig.~\ref{fig:ln(-lnR)}) in a single experiment only. This straight line always has  slope one, independent on the order $n$ of the interaction. Moreover, its ordinate axis intercept is a function of $\rho$~\footnote{In a recent experiment~\cite{Shimizu2}, Shimizu et al. investigate a single system, but vary $C_3$ by changing the surface density. In this case, it is in fact possible to extract information on the potential and to confirm Friedrich's theory on the low energy asymptote.}.

In conclusion, ground state $^3$He atoms allow experimental access to the high energy behavior of quantum reflection from the attractive potential they sense from an $\alpha$-quartz surface. 
We confirm the high energy asymptotic expression given by~(\ref{Eq:asimptota_n}) and show that it is determined by the non-retarded van der Waals potential only.
 Deviation of the experimental data from this asymptote shows that the interaction potential near the surface falls off steeper than a pure van der Waals potential. This is due to the influence of Casimir forces even at the distance of 30 \AA~above the surface.
Our analysis, based on the complete theory on QR and covering the entire energy range, shows excellent agreement with the experimental data for the potential coefficients $C_4=23.6$ eV\AA$^4$ and $l=(10 \pm 1)$ nm. 
The interaction potential compares well to the one calculated by Kunc et al. \cite{Kunc} in the vicinity of the potential minimum. Moreover, $l$ perfectly matches the transition wavelength from the electronic ground to the first excited state of He .       

The limited saturation rate of the detector and the relatively large value of $\sigma$ prevented us from measuring the reflection at energies $>10^{-3} V_{0}$. We are currently using a highly efficient mass spectrometer detector ~\cite{MDeKieviet:rsi2000}, in order to explore an even higher energy range for QR where the influence of the repulsive wall becomes visible.  
\acknowledgments{We are grateful to the Konrad-Adenauer-Stiftung fellowship for supporting the work of V.D.}

\end{document}